\documentstyle[preprint,prd,aps,eqsecnum,amssymb,epsf]{revtex} 
\tighten

\newcommand{\RR}{{\Bbb R}}

\newcommand{\kv}{{\bf k}}
\newcommand{\zerov}{{\bf 0}}

\newcommand{\OO}{{\cal O}}

\newcommand{\supp}{{\rm supp}\,}

\begin{document}

\title{Bounds on negative energy densities in flat spacetime}
\author{C.J. Fewster\thanks{Electronic address: cjf3@york.ac.uk} and
S.P. Eveson\thanks{Electronic address: spe1@york.ac.uk}}
\address{
Department of Mathematics, University of York, \\
Heslington, York YO1 5DD, United Kingdom}
\date{May 7, 1998}
\maketitle
                   
\begin{abstract}
We generalise results of Ford and Roman which place lower bounds --
known as quantum inequalities -- on the renormalised energy density of a
quantum field averaged against a choice of sampling function. Ford and
Roman derived their results for a specific non-compactly supported
sampling function; here we use a different argument to obtain quantum
inequalities for a class of smooth, even and non-negative sampling
functions which are either compactly supported or decay rapidly at
infinity. Our results hold in $d$-dimensional Minkowski space ($d\ge 2$)
for the free real scalar field of mass $m\ge 0$. We discuss various
features of our bounds in $2$ and $4$ dimensions. In particular, for
massless field theory in $2$-dimensional Minkowski space, we show that
our quantum inequality is weaker than Flanagan's optimal bound by a
factor of $\frac{3}{2}$.
\end{abstract}

\section{Introduction}

The stress-energy tensor $T_{ab}$ is said to obey the weak energy
condition (WEC) if $T_{ab}u^au^b\ge 0$ for all timelike vectors $u^a$.
This condition is obeyed by all known forms of classical matter, but is 
violated in quantum field theory~\cite{EGJ} in which the
renormalised energy density may become arbitrarily negative at points of
spacetime. If extended regions of large negative energy density occur in
nature, a variety of exotic phenomena might be possible, ranging from
violations of the second law of thermodynamics and cosmic censorship to
the creation of time machines and `warp drive'. Accordingly, it is
important to understand the extent to which the weak energy condition
may be violated. In a series of
papers~\cite{Ford,FRflat,PFsRW,PFsta,FPRsin}, 
Ford, Roman and Pfenning have studied quantum field theory in various
flat and curved spacetimes and established lower bounds (known
as {\em quantum inequalities}) on the time averaged energy densities
measured by observers.\footnote{Interestingly, it turns out that {\em
space} averaged energy densities in 4-dimensions are not bounded
below~\cite{Helf}.} The results have been employed to argue against the
possibility of traversable wormholes~\cite{FRworm}, warp
drive~\cite{PFwarp} and also to discuss the r\^ole of negative energy
densities in the process of black hole evaporation~\cite{FRblac}.

To be specific, consider a free real scalar field of mass $m$ in
$d$-dimensional Minkowski space, and let $T^{\rm ren}_{ab}$ denote the
renormalised stress-energy tensor. Quantum inequalities provide lower
bounds on the averaged expected energy density
\begin{equation}
\rho_{f,\psi} = \int_{-\infty}^\infty dt\,\langle 
T^{\rm ren}_{00}(t,\zerov) \rangle_\psi f(t)
\label{eq:rfp}
\end{equation}
measured by a stationary observer at the spatial origin $\zerov$, where
$f$ is a non-negative sampling function and the angle brackets denote
the expectation value in the quantum state $\psi$. The 
sampling function employed by Ford {\em et al.} is the Lorentzian
function peaked at $t=0$
\begin{equation}
L(t) = \frac{\tau}{\pi(t^2+\tau^2)},
\end{equation}
in which $\tau$ sets the timescale over which sampling occurs. With this
sampling function, Ford and Roman have shown~\cite{FRflat} that in four
dimensional Minkowski space the averaged energy density $\rho_{f,\psi}$
obeys the bound
\begin{equation}
\rho_{L,\psi} \ge -\frac{3}{32\pi^2\tau^4}G(2m\tau)
\label{eq:QIFR4}
\end{equation}
for all (sufficiently well-behaved) states $\psi$ and any $\tau>0$, where
the real-valued function $G$ is independent of $\psi$ and is 
positive and strictly decreasing on
$\RR^+$ with $G(0)=1$. In two
dimensions, they obtained the corresponding result
\begin{equation}
\rho_{L,\psi} \ge -\frac{1}{8\pi\tau^2}F(2m\tau)
\label{eq:QIFR2}
\end{equation}
where $F(y)=\frac{1}{2}y^2(K_0(y)+K_2(y))$ and the $K_\nu$ are modified
Bessel functions of the second kind. 

In the present paper, we will improve and generalise these bounds to
cover more general sampling functions in Minkowksi space of arbitrary
dimension. Some progress in this direction
has already been made by Flanagan~\cite{Flan} in the case of
2-dimensional massless field theory. Flanagan derived an
optimal lower bound on $\rho_{f,\psi}$ given by
\begin{equation}
\min_\psi \rho_{f,\psi} = -\frac{1}{24\pi}\int_{\supp f} dt
\,\frac{(f'(t))^2}{f(t)} 
\label{eq:QIFlan}
\end{equation}
for any smooth non-negative sampling function $f$, where $\supp f$
denotes the support of $f$.\footnote{Flanagan also derives related
quantum inequalities for other components of $T_{ab}^{\rm ren}$.}
Flanagan's argument depends critically on special features of
2-dimensional massless field theory. In this paper, we will consider
arbitrary smooth, non-negative, even
sampling functions $f$ of rapid decrease at infinity (including the
possibility of compact support), and will obtain the slightly weaker bound 
\begin{equation}
\rho_{f,\psi} \ge -\frac{1}{16\pi}\int_{\supp f} dt
\,\frac{(f'(t))^2}{f(t)} 
\end{equation}
in the 2-dimensional massless case. However, our argument has the virtue of
generalising directly to both massive field theory in two dimensions and
to massive and massless field theory in four dimensional Minkowski
space. Our general quantum inequality in $n+1$-dimensional
Minkowski space, is the bound
\begin{equation}
\rho_{f,\psi} \ge
-\frac{C_n}{2\pi(n+1)}
\int_m^\infty du\, \left(\widehat{f^{1/2}}(u)\right)^2 u^{n+1}
Q_n\left(\frac{u}{m}\right),
\label{eq:QI}
\end{equation}
for all (sufficiently well-behaved) quantum states $\psi$, where
$f^{1/2}(t)=\sqrt{f(t)}$, the hat denotes the Fourier transform and the  
constant $C_n$ is equal to the area of the unit $n-1$-sphere divided by
$(2\pi)^n$. The bounded non-negative functions $Q_n$ are defined (for
$n\ge 1$) on $[1,\infty)$ by
\begin{equation}
Q_n(x) = (n+1)x^{-(n+1)}\int_1^x dy\, y^2(y^2-1)^{n/2-1},
\label{eq:Qn}
\end{equation}
and obey $Q_n(1)=0$ and $Q_n(x)\to 1$ as $x\to\infty$.
We plot $Q_1$ and $Q_3$ in Figures~1 and~2. 

The derivation of Eq.~(\ref{eq:QI}) is quite simple and depends mainly
on the canonical commutation relations and the convolution theorem.
Although we will argue formally here, we expect that the elementary nature 
of the argument will facilitate a fully rigorous treatment.

\section{Preliminaries}

We begin by stating our conventions.
In $n+1$-dimensional Minkowski space, $\eta_{ab}$ denotes the metric
with signature $+-\cdots -$. The Klein--Gordon equation is
\begin{equation}
(\square +m^2)\varphi \equiv 
\left(\eta^{ab}\partial_a\partial_b+m^2\right)\varphi
= 0
\end{equation}
and the quantum field $\Phi(x)$ is defined by
\begin{equation}
\Phi(x) = \int\frac{d^n\kv}{(2\pi)^n\sqrt{2\omega(\kv)}} 
\left( a(\kv) e^{-ik\cdot x} +  a^\dagger(\kv) e^{ik\cdot x}
\right)
\end{equation}
where $\omega(\kv)=\sqrt{\|\kv\|^2+m^2}$, 
the $n+1$-vector $k^a$ has components $(\omega(\kv),\kv)$, 
and the annihilation and creation
operators $a(\kv)$ and $a^\dagger(\kv)$ obey the canonical commutation 
relations 
\begin{equation}
[a(\kv),a(\kv')]=[a^\dagger(\kv),a^\dagger(\kv')]=0 \qquad
[a(\kv),a^\dagger(\kv')]=(2\pi)^n\delta(\kv-\kv').
\end{equation}
The classical energy density of a
field $\varphi$ is
\begin{equation}
T_{00} = \frac{1}{2}\left\{\left(\partial_0 \varphi\right)^2+
\sum_{i=1}^n\left(\partial_i \varphi\right)^2 +m^2\phi^2\right\},
\end{equation}
from which the renormalised (normal ordered) quantum energy density
at position $(t,\zerov)$ is easily shown to be
\begin{eqnarray}
T^{\rm ren}_{00}(t,\zerov)&=&\int
\frac{d^n\kv\,d^n\kv'}{(2\pi)^{2n}}
\left[\frac{\omega(\kv)\omega(\kv')+\kv\cdot\kv'}{4(\omega(\kv)\omega(\kv'))^{1/2}}
\left\{a^\dagger(\kv)a(\kv')e^{i(\omega(\kv)-\omega(\kv'))t}
-a(\kv)a(\kv')e^{-i(\omega(\kv)+\omega(\kv'))t}
\right\} \right.\nonumber\\
&&+\left.\frac{m^2}{4(\omega(\kv)\omega(\kv'))^{1/2}}
\left\{a^\dagger(\kv)a(\kv')e^{i(\omega(\kv)-\omega(\kv'))t}
+a(\kv)a(\kv')e^{-i(\omega(\kv)+\omega(\kv'))t} 
\right\} +{\rm h.c.}
\right].
\label{eq:T00}
\end{eqnarray}

Finally, the Fourier transform $\widehat{f}$ of a function $f$ on $\RR$
is defined by
\begin{equation}
\widehat{f}(\omega)= \int_{-\infty}^\infty dt\, f(t)e^{-i\omega t}.
\end{equation}

\section{A positivity result}\label{sect:pos}

Let $f$ be a smooth, even and non-negative function on $\RR$ which decays
rapidly at infinity [including the possibility that $f$ is compactly
supported]. Using $f^{1/2}$ to denote the pointwise square root of $f$
[i.e., $f^{1/2}(t)=\sqrt{f(t)}$], the function $g$ defined by
\begin{equation}
g(\omega) = \frac{\widehat{f^{1/2}}(\omega)}{\sqrt{2\pi}}
\end{equation}
on $\RR$ is smooth, real-valued and even, decays rapidly at infinity and
obeys
\begin{equation}
g\star g =\widehat{f},
\end{equation}
where the convolution $\star$ is defined by 
\begin{equation}
(h_1\star h_2)(\omega) = \int_{-\infty}^\infty d\omega'\,
h_1(\omega-\omega') h_2(\omega').
\end{equation}

Now let $p$ be a real valued function on $\RR^n$, growing no faster than
polynomially. We use the annihilation and creation
operators of the scalar field in $n+1$-dimensional Minkowski space to
define two families $\{\OO_\omega^\pm\mid \omega\in\RR^+\}$ of operators
on the Fock space of the Minkowski vacuum by
\begin{equation}
\OO_\omega^\pm = \int\frac{d^n\kv}{(2\pi)^n}
\left\{ g(\omega-\omega(\kv))a(\kv)\pm
g(\omega+\omega(\kv))a^\dagger(\kv)\right\}
p(\kv).
\end{equation}
Using the commutation relations and symmetrising the integrand in $\kv$
and $\kv'$, we calculate
\begin{eqnarray}
\int_0^\infty \!d\omega\, {\OO^\pm_\omega}^\dagger\OO^\pm_\omega &=&
\int_0^\infty d\omega\int\frac{d^n\kv\,d^n\kv'}{(2\pi)^{2n}}
\left\{ 
g(\omega-\omega(\kv))g(\omega-\omega(\kv'))a^\dagger(\kv)a(\kv')\right.
\nonumber \\
&&+ g(\omega+\omega(\kv))g(\omega+\omega(\kv'))a(\kv)a^\dagger(\kv')
\nonumber \\
&& \pm 
g(\omega-\omega(\kv))g(\omega+\omega(\kv'))a^\dagger(\kv)a^\dagger(\kv')
\nonumber \\
&& \pm\left. g(\omega+\omega(\kv))g(\omega-\omega(\kv'))a(\kv)a(\kv')
\right\}
p(\kv)p(\kv')
\nonumber \\
&=& S^\pm +\int_0^\infty d\omega\int\frac{d^n\kv}{(2\pi)^{n}}
g(\omega+\omega(\kv))^2p(\kv)^2,
\label{eq:OO}
\end{eqnarray}
where
\begin{eqnarray}
S^\pm&=& \frac{1}{2}\int\frac{d^n\kv\,d^n\kv'}{(2\pi)^{2n}}
\left\{
F(\kv,\kv')\left(a^\dagger(\kv)a(\kv')+a^\dagger(\kv')a(\kv)\right)\right.
\nonumber\\
&&\pm\left.
G(\kv,\kv')\left(a^\dagger(\kv)a^\dagger(\kv')+a(\kv')a(\kv)\right)
\right\}p(\kv)p(\kv')
\end{eqnarray}
and the functions $F$ and $G$ are given by
\begin{equation}
F(\kv,\kv')=\int_0^\infty d\omega\,
g(\omega-\omega(\kv))g(\omega-\omega(\kv'))+ 
g(\omega+\omega(\kv))g(\omega+\omega(\kv'))
\end{equation}
and
\begin{equation}
G(\kv,\kv')=\int_0^\infty d\omega\,
g(\omega-\omega(\kv))g(\omega+\omega(\kv'))+ 
g(\omega+\omega(\kv))g(\omega-\omega(\kv')).
\end{equation}
The expressions for $F$ and $G$ may be simplified, using the fact that
$g$ is even, to obtain
\begin{eqnarray}
F(\kv,\kv')&=&\int_{-\infty}^\infty d\omega
g(\omega-\omega(\kv))g(\omega-\omega(\kv')) \nonumber\\
&=& (g\star g)(\omega(\kv)-\omega(\kv'))\nonumber \\
&=& \widehat{f}(\omega(\kv)-\omega(\kv'))
\label{eq:Fkk}
\end{eqnarray}
and, similarly,
\begin{equation}
G(\kv,\kv')=(g\star g)(\omega(\kv)+\omega(\kv'))=
\widehat{f}(\omega(\kv)+\omega(\kv')).
\label{eq:Gkk}
\end{equation}

Since the right hand side of Eq.~(\ref{eq:OO}) is (formally) a manifestly
positive operator, we conclude that the expectation value $\langle
S^\pm\rangle_\psi$ 
obeys the following bound
\begin{eqnarray}
\langle S^\pm \rangle_\psi &\ge& -\int_0^\infty
d\omega\int\frac{d^n\kv}{(2\pi)^{n}}
g(\omega+\omega(\kv))^2p(\kv)^2 \nonumber \\
&=& -\frac{1}{2\pi}\int_0^\infty
d\omega\int\frac{d^n\kv}{(2\pi)^{n}}
\widehat{f^{1/2}}(\omega+\omega(\kv))^2p(\kv)^2
\label{eq:bd}
\end{eqnarray}
in all sufficiently well behaved states $\psi$. As we will see, this
bound provides the key to our derivation of the quantum
inequality~(\ref{eq:QI}).  

We conclude this section with two remarks. Firstly, we have argued
rather formally and have interchanged orders of integration at will.
Nonetheless, we expect that if $f$ has rapid decay at infinity (e.g., if
$f$ is a Schwartz test function) our result can be established rigorously
for a dense set of states in the folium of the usual Minkowski vacuum,
and conceivably for all Hadamard states on the usual field algebra.
Secondly, we do not claim that Eq.~(\ref{eq:bd}) is the sharpest bound
that can be placed on the expectation value of $S^\pm$, and in fact do not
expect that the bound is actually attained by any reasonable state
$\psi$, as $\psi$ would necessarily belong to the kernel of all
$\OO_\omega^\pm$'s (except, perhaps, for a set of measure zero).

\section{Derivation of the quantum inequalities}

The result derived above allows a simple proof of the energy
inequalities. With $f$ as in Section~\ref{sect:pos}, define
\begin{equation}
T_f = \int_{-\infty}^\infty T_{00}^{\rm ren}(t,\zerov) f(t)
\end{equation}
so that $\rho_{f,\psi} = \langle T_f\rangle_\psi$. By Eq.~(\ref{eq:T00})
and the fact that $\widehat{f}$ is even, we have
\begin{eqnarray}
T_f &=& \frac{1}{2}\int\frac{d^n\kv\,d^n\kv'}{(2\pi)^{2n}}
\frac{\omega(\kv)\omega(\kv')+\kv\cdot\kv'}{2(\omega(\kv)\omega(\kv'))^{1/2}}
\left\{F(\kv,\kv')
\left(a^\dagger(\kv)a(\kv')+a^\dagger(\kv')a(\kv)\right)\right.
\nonumber\\
&&+\left.
G(\kv,\kv')\left(a^\dagger(\kv)a^\dagger(\kv')+a(\kv')a(\kv)\right)
\right\} \nonumber\\
&&+\frac{1}{2}\int\frac{d^n\kv\,d^n\kv'}{(2\pi)^{2n}}
\frac{m^2}{2(\omega(\kv)\omega(\kv'))^{1/2}}
\left\{F(\kv,\kv')
\left(a^\dagger(\kv)a(\kv')+a^\dagger(\kv')a(\kv)\right)\right.
\nonumber\\
&&-\left.
G(\kv,\kv')\left(a^\dagger(\kv)a^\dagger(\kv')+a(\kv')a(\kv)\right)
\right\}
\end{eqnarray}
with $F$ and $G$ given by Eqs.~(\ref{eq:Fkk}) and~(\ref{eq:Gkk}).
Clearly, $T_f$ is a finite sum of operators of the form $S^\pm$.
We therefore apply the
bound~(\ref{eq:bd}) for the various cases 
\begin{equation}
p(\kv)=\frac{\omega(\kv)}{\sqrt{2\omega(\kv)}},\qquad
p(\kv)=\frac{k_i}{\sqrt{2\omega(\kv)}},\qquad
p(\kv)=\frac{m}{\sqrt{2\omega(\kv)}}
\end{equation}
and add the results, obtaining
\begin{eqnarray}
\rho_{f,\psi} &\ge& -\frac{1}{2\pi}
\int_0^\infty
d\omega\int\frac{d^n\kv}{(2\pi)^{n}}\omega(\kv)
\widehat{f^{1/2}}(\omega+\omega(\kv))^2 \nonumber \\
&=&-\frac{C_n}{2\pi}
\int_0^\infty d\omega\int_m^\infty d\omega'\,
\widehat{f^{1/2}}(\omega+\omega')^2{\omega'}^2\left({\omega'}^2-m^2\right)^{n/2-1}
\end{eqnarray}
for all (sufficiently well-behaved) states $\psi$.
Here, $C_n$ is equal to the area of the unit $n-1$-sphere divided by
$(2\pi)^n$ [with the convention that $C_1=1/\pi$], that is,
\begin{equation}
C_n = \frac{1}{2^{n-1}\pi^{n/2}\Gamma(\frac{1}{2}n)}.
\end{equation}

\begin{figure}
\center{\leavevmode\epsfxsize=6truein\epsfbox{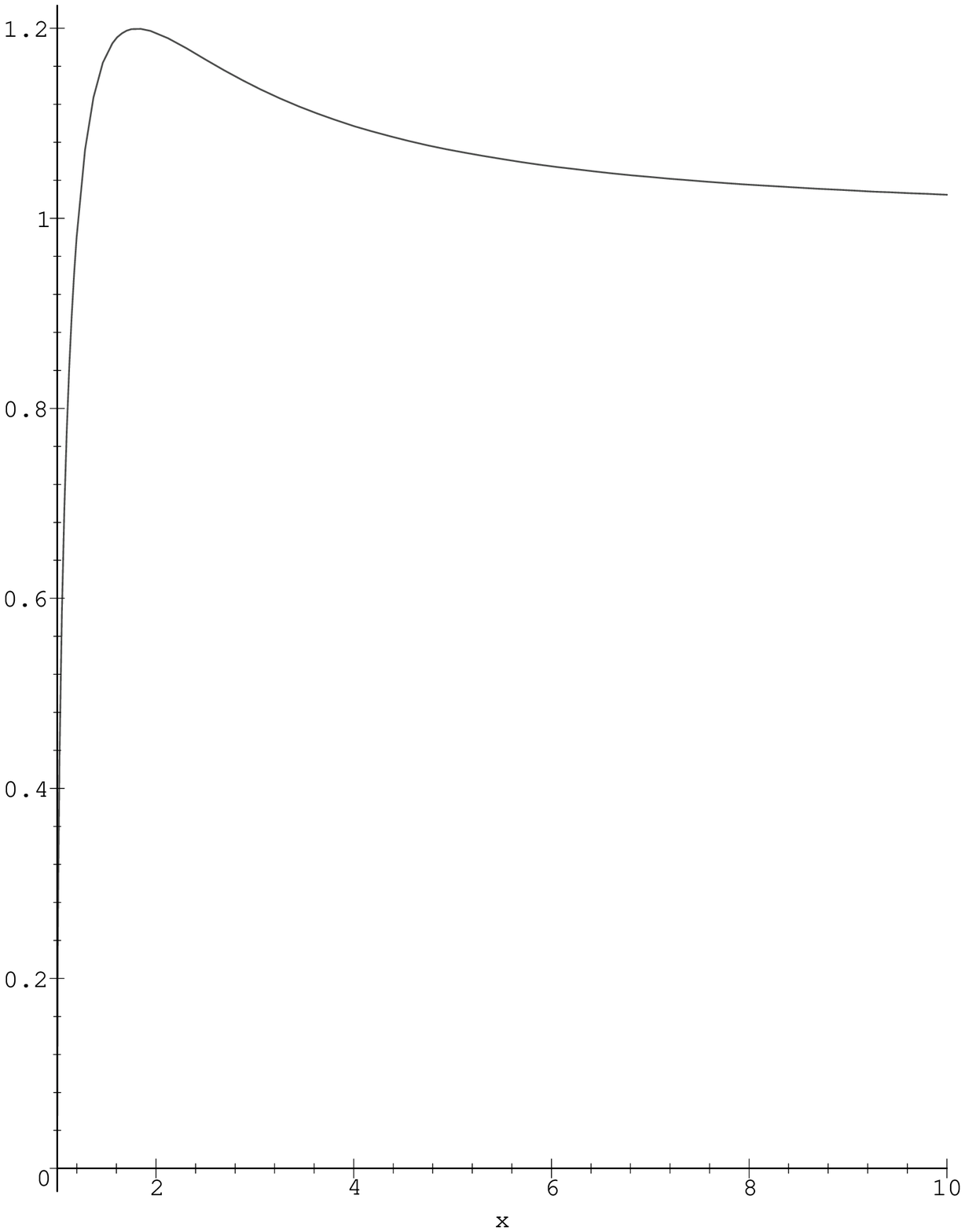}}
\caption{Graph of $Q_1(x)$.}
\label{fig1}
\end{figure}

\begin{figure}
\center{\leavevmode\epsfxsize=6truein\epsfbox{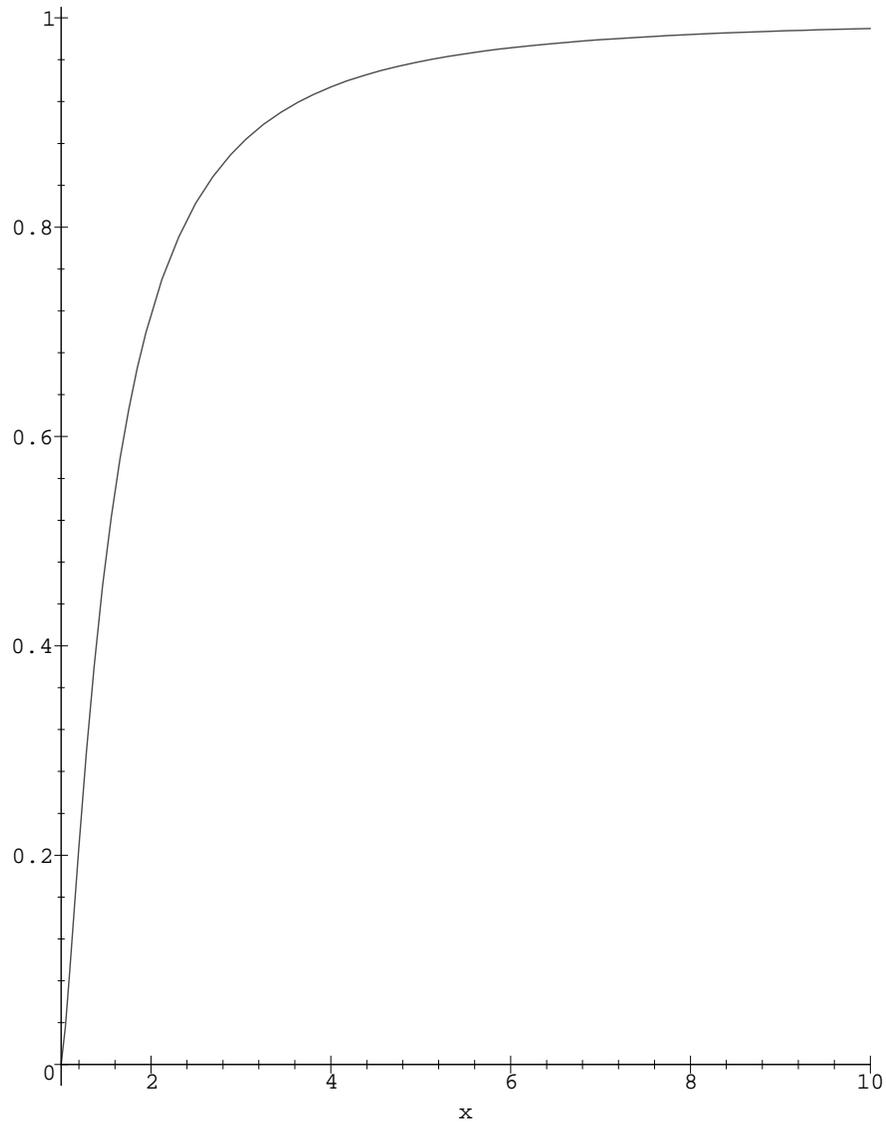}}
\caption{Graph of $Q_3(x)$.}
\label{fig2}
\end{figure}

If we now make the change of variables 
\begin{equation}
u=\omega+\omega' \qquad v=\omega'
\end{equation}
we find
\begin{eqnarray}
\rho_{f,\psi} &\ge& -\frac{C_n}{2\pi}
\int_m^\infty du\, \left(\widehat{f^{1/2}}(u)\right)^2
\int_m^u dv\, v^2(v^2-m^2)^{n/2-1} \nonumber \\
&=&  -\frac{C_n}{2\pi(n+1)}
\int_m^\infty du\, \left(\widehat{f^{1/2}}(u)\right)^2 u^{n+1}
Q_n\left(\frac{u}{m}\right),
\end{eqnarray}
where the functions $Q_n(x)$ are defined by Eq.~(\ref{eq:Qn}), thus
completing the derivation of our general quantum inequality
Eq.~(\ref{eq:QI}). We note that each $Q_n(x)$ is a positive function 
with $Q_n(1)=0$ and $Q_n(x)\to 1$
as $x\to\infty$. For $n\ge 2$, $Q_n(x)$ is strictly increasing on
$\RR^+$, while $Q_1(x)$ exhibits a maximum near $x=1.8$ and decreases
thereafter -- see Figures~\ref{fig1} and~\ref{fig2}.

To conclude this section, we consider the scaling behaviour
of~(\ref{eq:QI}). Let $f_\lambda(t)$ be the scaled function
\begin{equation}
f_\lambda(t) = \lambda^{-1} f(t/\lambda).
\end{equation}
This function has the same integral over $\RR$ as $f$ but with $\lambda$
times the characteristic width. It is easy to see that
\begin{equation}
\left(\widehat{f_{\lambda}^{1/2}}(u)\right)^2 = \lambda 
\left(\widehat{f^{1/2}}(u)\right)^2
\end{equation}
from which it follows that the bound for sampling function $f_\lambda$
at mass $m$ is equal to $\lambda^{-(n+1)}$ times the bound for sampling
function $f$ at mass $\lambda m$. This is the expected scaling
behaviour and is also exhibited by the quantum inequalities~(\ref{eq:QIFR4})
and~(\ref{eq:QIFR2}) of Ford and Roman. Furthermore, it is clear that
the value of the bound~(\ref{eq:QI}) tends to zero for each fixed $f$ as
$m\to \infty$. Thus we have
\begin{equation}
\lim_{\lambda\to\infty} \lambda \rho_{f_\lambda,\psi} \ge 0
\end{equation}
and we recover the averaged weak energy condition in the limit
(cf.~\cite{FRavg}). 

\section{Special cases}

In this section, we briefly discuss the most interesting cases of the
general quantum inequality derived above, and compare our results with
those of Flanagan~\cite{Flan} and Ford and Roman~\cite{FRflat}. 

\subsection{Two dimensions}

As mentioned in the introduction, Flanagan has derived an apparently
optimal quantum inequality for massless 2-dimensional field
theory~\cite{Flan}. Substituting $n=1$ and $m=0$ into our bound 
Eq.~(\ref{eq:QI})
and using the fact that $Q_1(x)\to 1$ as $x\to\infty$, we find
\begin{equation}
\rho_{f,\psi} \ge -\frac{1}{4\pi^2}\int_0^\infty du\, u^2
\left(\widehat{f^{1/2}}(u)\right)^2.
\end{equation}
The integrand is an even function in $u$, so we may extend the range of
integration to the whole of $\RR$ and then employ Parseval's theorem to
yield a $t$-space version of the quantum inequality
\begin{equation}
\rho_{f,\psi} \ge -\frac{1}{4\pi}\int_{-\infty}^\infty 
dt\, \left({f^{1/2}}'(t)\right)^2 =
-\frac{1}{16\pi}\int_{{\rm supp}\,f} dt\, \frac{\left(f'(t)\right)^2}{f(t)},
\end{equation}
which should be compared with Flanagan's bound~(\ref{eq:QIFlan}). Our
bound is seen to be weaker by a factor of $\frac{3}{2}$; there is no
contradiction because we do not expect our bound to be
optimal. Since Flanagan's bound is six times stronger than that of Ford
and Roman when applied to the Lorentzian sampling function, our bound is
accordingly four times stronger in this case. 

In the 2-dimensional massive case, the quantum inequality~(\ref{eq:QI})
becomes
\begin{equation}
\rho_{f,\psi}\ge -\frac{1}{4\pi^2}\int_0^\infty du\, u^2
\left(\widehat{f^{1/2}}(u)\right)^2 Q_1\left(\frac{u}{m}\right)
\end{equation}
with $Q_1$ given by
\begin{equation}
Q_1(x) = \left(1-x^{-2}\right)^{1/2} + x^{-2}\log(x+(x^2-1)^{1/2}).
\end{equation}
Figure~\ref{fig1} shows that $Q_1(x)$ exhibits a maximum value of
approximately $1.2$ near $x=1.8$. We cannot exclude the possibility that
massive fields in $2$-dimensions can exhibit slightly stronger negative
energy densities than massless fields can (by a factor of at most $1.2$).
Interestingly, a similar phenomenon occurs in Ford and Roman's
treatment~\cite{FRflat} so it would be worthwhile to determine
whether this is indeed so, or whether the peak is an artifact of the
argument (as Ford and Roman suggest). 

\subsection{Four dimensions}

Just as in the 2-dimensional case discussed above, the 4-dimensional
quantum inequality takes a particularly simple form for massless fields:
\begin{equation}
\rho_{f,\psi} \ge -\frac{1}{16\pi^3}\int_{0}^\infty du\,
\left(\widehat{f^{1/2}}(u)\right)^2 u^4 = 
 -\frac{1}{16\pi^2}\int_{-\infty}^\infty dt\,\left(
{f^{1/2}}''(t)\right)^2.
\label{eq:QI4massless}
\end{equation}
In particular, for the Lorentzian function we have
\begin{equation}
\rho_{L,\psi}\ge -\frac{1}{16\pi^2} \int_{-\infty}^\infty dt\,
\frac{(2t^2-\tau^2)^2\tau}{\pi(t^2+\tau^2)^5}
=-\frac{27}{2048\pi^2\tau^4}
\end{equation}
which is $\frac{9}{64}$ of Ford and Roman's
result~(\ref{eq:QIFR4}) in this case [recall that $G(0)=1$]. This entails
a slight tightening of the constraints on traversable
wormholes~\cite{FRworm}. 

Finally, we state the form of Eq.~(\ref{eq:QI}) in the 4-dimensional
massive case. In terms of the Fourier transform of $f$, we have
\begin{equation}
\rho_{f,\psi} \ge -\frac{1}{16\pi^3}\int_m^\infty du\,\left(
\widehat{f^{1/2}}(u)\right)^2 u^4 Q_3\left(\frac{u}{m}\right)
\end{equation}
where
\begin{equation}
Q_3(x) =\left(1-x^{-2}\right)^{1/2}\left(1-\frac{1}{2x^2}\right)
-\frac{1}{2x^4}\log\left(x+(x^2-1)^{1/2}\right).
\end{equation}
As Figure~\ref{fig2} makes clear, the function $Q_3$ is bounded between
zero and unity. Accordingly the bound~(\ref{eq:QI4massless}) is also a
lower bound for massive fields: in 4 dimensions [indeed, in any
spacetime dimension greater than or equal to three] the effect of
introducing a mass cannot decrease the averaged energy density below the
massless bound. 

\section{Conclusion}

We have given a simple derivation of a quantum inequality for
the free real scalar field in Minkowski space of any dimension, which
allows more general
sampling functions than previously possible. In particular, our
derivation allows for compactly supported sampling functions and
therefore removes any remaining doubt that the quantum inequalities of
Ford and Roman might rely on subtle large scale effects to cancel local
negative energy densities. In conclusion, we make various remarks. 

First, the derivation given here has been somewhat formal and lacking
in mathematical rigour. However, we hope that our argument is simple
enough that a rigorous formulation might be established without too much
difficulty, and intend to return to this issue elsewhere. Two elements
of our discussion need to be more precisely specified: namely the class of
sampling functions and the class of quantum states for which
Eq.~(\ref{eq:QI}) is valid. We have required that the sampling function
$f$ be even primarily for convenience, and expect that this condition may
be removed. In addition, we may speculate
as to whether it is necessary that $f$ be smooth. Of course, a necessary
condition on $f$ is is that
the integral on the right hand side of Eq.~(\ref{eq:QI}) should
converge; this amounts to a smoothness condition on $f$ which becomes
more stringent as the spacetime dimension increases (and which is
always satisfied if $f$ is actually smooth). Thus, it may be that the
smoothness of $f$ could be relaxed to $C^k$ where $k$ depends on $n$. 
However, it is clearly important that $f$ has some degree of continuity.
As an example, suppose $f$ is the characteristic function for the
interval $[-\tau,\tau]\subset\RR$. We have 
\begin{equation}
\left(\widehat{f^{1/2}}(\omega)\right)^2=\frac{4\sin^2
\omega\tau}{\omega^2}, 
\end{equation} 
from which it follows that the integral in Eq.~(\ref{eq:QI}) diverges
for any $n\ge 1$. Thus the quantum inequality provides no information in
this case, which is consistent with results of Garfinkle (quoted by
Yurtsever in~\cite{Yurt} -- see particularly footnote~[1] therein) that the
integral of $T^{\rm ren}_{00}$ over sharply defined boxes in spacetime
can be unboundedly negative. As mentioned above, the class of quantum
states for which Eq.~(\ref{eq:QI}) holds must also be clarified. It is
likely that quantum inequalities will hold for a dense class of states
in the Fock space built on the Minkowski vacuum; more generally, we hope
that such inequalities might be established for the class of (globally)
Hadamard states~\cite{KW,Rad}.

Second, we have seen that our bound is weaker by a factor of
$\frac{3}{2}$ than
the optimal bound proposed by Flanagan~\cite{Flan}, which was derived
using special 
features of 2-dimensional massless field theory. It would be interesting
to investigate whether our argument could be improved to replicate
Flanagan's result and perhaps to obtain optimal bounds for the massive
case and also higher dimensional spacetimes.

Finally, we anticipate no particular difficulties in generalising our
argument to provide quantum inequalities in curved spacetime. Again, we
intend to return to this elsewhere.

\widetext

\widetext

\end{document}